\documentclass[iop,apjl]{emulateapj}
\usepackage[varg]{txfonts}

\newcommand{\beq}{\begin{equation}}
\newcommand{\eeq}{\end{equation}}
\newcommand{\beqn}{\begin{eqnarray}}
\newcommand{\eeqn}{\end{eqnarray}}
\renewcommand{\leq}{\leqslant}

\newcommand{\eqref}[1]{(\ref{#1})}

\newcommand{\AU}{{\rm AU}} 

\newcommand{\md}{{\rm mid}}
\newcommand{\ideal}{{\rm ideal}}
\newcommand{\res}{{\rm res}}

\newcommand{\alphacore}{\alpha_{\rm core}}

\newcommand{\nvf}{{\rm NVF}}

\defcitealias{O09}{O09}
\defcitealias{OH11}{OH11}

\shorttitle{PLANETESIMAL FORMATION IN DEAD ZONES}
\shortauthors{OKUZUMI \& HIROSE}
\slugcomment{ApJL Accepted, May 30, 2012}

\begin{document}
\title{Planetesimal Formation in Magnetorotationally Dead Zones:
Critical Dependence on the Net Vertical Magnetic Flux}

\author{Satoshi Okuzumi}
\affil{Department of Physics, Nagoya University, Nagoya, Aichi 464-8602, Japan; okuzumi@nagoya-u.jp}
\and
\author{Shigenobu Hirose}
\affil{Institute for Research on Earth Evolution, JAMSTEC, Yokohama, Kanagawa 236-0001, Japan}

\begin{abstract}
Turbulence driven by magnetorotational instability (MRI) affects planetesimal formation by inducing diffusion and collisional fragmentation of dust particles. 
We examine conditions preferred for planetesimal formation in MRI-inactive ``dead zones'' using an analytic dead-zone model based on our recent resistive MHD simulations. We argue that successful planetesimal formation requires not only a sufficiently large dead zone (which can be produced by tiny dust grains) but also a sufficiently small net vertical magnetic flux (NVF).  Although often ignored, the latter condition is indeed important since the NVF strength determines the saturation level of turbulence in MRI-active layers. We show that direct collisional formation of icy planetesimal across the fragmentation barrier is possible when the NVF strength is lower than 10~mG (for the minimum-mass solar nebula model). Formation of rocky planetesimals via the secular gravitational instability is also  possible within a similar range of the NVF strength. Our results indicate that the fate of planet formation largely depends on how the NVF is radially transported in the initial disk formation and subsequent disk accretion processes.
 \end{abstract}
\maketitle

\section{Introduction}
Formation of kilometer-sized planetesimals is the initial step of planet formation 
in protoplanetary disks.
Several mechanisms have been proposed for planetesimal formation, which include 
gravitational instability (GI) of dust subdisks \citep[e.g.,][]{GW73,J+07,Y11}
and direct coagulation \citep[e.g.,][]{WC93,OTKW12}. 
However, the outcome of these processes strongly depends on
the turbulent state of the gas disk. 
Turbulence is know to concentrate particles of particular sizes,
which could assist their gravitational collapse \citep{C+01,J+07}.
On the other hand, turbulence also stirs up dust subdisks 
and thereby stabilizes GI \citep{TCS10}.
In addition, the relative velocity induced by turbulence 
can lead to catastrophic disruption of large dust particles \citep{J+08}.

The magnetorotational instability (MRI; \citealt*{BH91}) has been 
thought as the most plausible driving mechanism of protoplanetary disk turbulence.
The activity of MRI largely depends on non-ideal MHD effects \citep[e.g.,][]{SM99,BS11,WS12}.
A high ohmic resistivity creates an MRI-inactive ``dead zone'' near the midplane \citep{G96}, which reduces turbulence strength. 
Importantly, the size of the dead zone depends on the amount of tiny dust particles,
because they effectively 
reduce the gas ionization degree and hence enhance the resistivity \citep{SMUN00}.
Although the dead zone has often been invoked as 
a favorable site for planetesimal formation \citep[e.g.,][]{C07,BS10,Y11}, 
self-consistent modeling of the MRI--dust coevolution has not been done so far.

Another important, but much less appreciated factor is the {\it net vertical flux} (NVF) 
of the magnetic fields. The NVF indeed matters since it determines the saturation 
level of MRI-driven turbulence \citep[e.g.,][]{HGB95,SMI10}.
This is especially true when a large dead zone is present, for which case
the vertically integrated accretion stress decreases with decreasing 
NVF (\citealt*{GNT12,OH11}, henceforth \citetalias{OH11}).

In this Letter, we investigate possible pathways of planetesimal formation 
taking into account the dependence of MRI-driven turbulence strength
on dust size distribution and NVF strength.
In our previous paper \citepalias{OH11}, we systematically studied 
the saturated state of MRI-driven turbulence with local stratified, ohmic-resistive MHD simulations. 
We obtained an analytic prescription for the saturation level as 
a function of the ohmic diffusivity and NVF strength.
Using this prescription together with ionization balance calculation including grain charging, 
we are able to determine turbulence strength consistently with 
the amount of tiny grains and NVF strength.
We consider the formation of rocky and icy planetesimals independently, 
because icy particles have a high sticking efficiency compared to rocky particles \citep{CTH93,GKBB11}.
We test direct collisional formation of icy planetesimals outside the snow line,
and rocky planetesimal formation via GI inside the snow line.

\section{Model Description}\label{sec:model}
We consider a protoplanetary disk around a solar mass star.
We take the gas surface density $\Sigma_g$ from the minimum-mass solar nebula 
(MMSN) model of \citet{H81}.
The gas temperature is taken from the passive, optically thick disk model of \citet{CG97}. 
This model well approximates the gas temperature inside a dead zone,  
since MRI-driven heating occurs mainly at the upper boundary of the active layer 
where the optical depth is small \citep{HT11}. 
The assumed temperature gives a snow line at orbital radius $r\sim1~\AU$.
The gas density $\rho_g$ depends on the distance $z$ from the midplane as 
$\rho_g=(\Sigma_g/\sqrt{2\pi} h)\exp(-z^2/2h^2)$, where $h$ is the gas scale height
given as the sound speed $c_s$ divided by the Keplerian frequency $\Omega$.
Because of gas pressure support, the gas disk rotates at a slightly sub-Keplerian velocity.
This causes systematic radial drift of dust particles relative to the mean gas motion. 
The drift speed reaches the maximum $|v_{dr,\max}|\approx30~{\rm m~s^{-1}}$ 
when the stopping time $t_s$ of the particle equals $\Omega^{-1}$ \citep{W77}.
At $r\sim1$--$5~\AU$, the dimensionless stopping time $\Omega t_s=1$ 
corresponds to particle radius $a\sim 1~{\rm m}$.

We consider MRI-driven with a dead zone.
The most important turbulent quantity for dust evolution is  
the random velocity of the gas since it 
determines the collision velocity
and diffusion coefficient 
of dust particles. 
\citetalias{OH11} found that the gas velocity dispersion at the midplane, 
$\delta v_{g,\md}$, is well approximated as 
\beq
\delta v_{g,\md}=\sqrt{0.78\alphacore}c_s, 
\label{eq:dvmid}
\eeq
where $\alphacore$ is the  accretion stress integrated over low altitudes and 
normalized by $\Sigma_g c_s^2$.
Equation~\eqref{eq:dvmid} holds no matter if the midplane is magnetically dead,
since hydrodynamical waves created in active layers propagate
across dead zone boundaries. 
\citetalias{OH11} also found that $\alphacore$ is 
determined by the dead zone size and NVF as
\beq
\alphacore = \frac{510}{\beta_{\nvf,\md}}\exp\left(-\frac{0.54h_{\res}}{h}\right)
+0.011\exp\left(-\frac{3.6h_{\Lambda}}{h}\right),
\label{eq:pred_core}
\eeq
where $\beta_{\nvf,\md} = 8\pi\rho_{g,\md}c_s^2/B_\nvf^2$ 
is the midplane plasma beta defined by the NVF strength $B_\nvf$, and 
$h_{\rm res}$ and $h_\Lambda$ are quantities that characterize 
the vertical extent of the dead zone.
Precisely, $h_{\rm res}$ is the height below which 
the characteristic MRI wavelength $\lambda_\res=2\pi\eta/v_{Az}$ 
in the resistive MHD limit exceeds $h$, 
whereas $h_\Lambda$ is the height below which the ohmic Elsasser number 
$\Lambda = v_{Az}^2/\eta\Omega$ falls below unity, 
where $v_{Az}=B_{\nvf}/\sqrt{4\pi\rho_g}$ is the Alfv\'{e}n speed defined 
by the NVF and $\eta$ is the ohmic diffusivity. 
Linear stability analysis \citep{SM99} shows that ohmic resistivity suppresses 
the most unstable MRI mode at $\Lambda\la 1$ ($z\la h_\Lambda$) 
but less unstable modes survive as long as $\lambda_\res\la h$ ($z \ga h_\res$). 
Thus, the region $h_\res<z<h_\Lambda$ can be interpreted 
as the transition layer between the active and dead zones.
The active layer extends up to $z=h_\ideal$, above which
MRI is stabilized because of a low local plasma beta 
(\citealt*{SM99}; \citetalias{OH11}).
In the saturated state, $h_\ideal$ is approximately given by 
$h_\ideal = [2\ln(\beta_{\nvf,\md}/3000)]^{1/2}h$ 
according to the simulations of \citetalias{OH11}.

For given $\delta v_{g,\md}$, 
the gas random velocity $\delta v_g$ at each height $z$
is given as $\delta v_g=\exp(z^2/4h^2)\delta v_{g,\md}$ \citepalias{OH11}. 
The vertical diffusion coefficient for dust is given by 
$D_z=D_{z0}/[1+(\Omega t_s)^2]$ \citep{YL07}, where 
\beq
D_{z0}\approx\delta v_g^2/3\Omega
\label{eq:Dz0}
\eeq
is the vertical diffusion coefficient for passive ($\Omega t_s\ll1$) 
contaminants (\citealt*{FP06}; \citetalias{OH11}). 
Note that Equation~\eqref{eq:Dz0} applies even in dead zones,
which is consistent with the fact that 
hydrodynamical waves propagating from active layers have a finite correlation
time $\sim\Omega^{-1}$ \citep{GNT12}.
The turbulence-driven relative velocity $\Delta v_t$ of dust particles
is calculated from the prescription of \citet{OC07} for Kolmogorov turbulence.
In reality, a Kolmogorov energy cascade may not be established 
for waves in dead zones.
However, the assumed $\Delta v_t$ at least gives a reasonable estimate 
of the relative velocity for marginally or fully decoupled ($\Omega t_s \ga1$) particles
since their relative velocity is determined by the largest-scale gas motion
with correlation time $\sim\Omega^{-1}$ \citep[see][]{OC07}.
 
\begin{figure}[t]
\epsscale{1.1}
\plotone{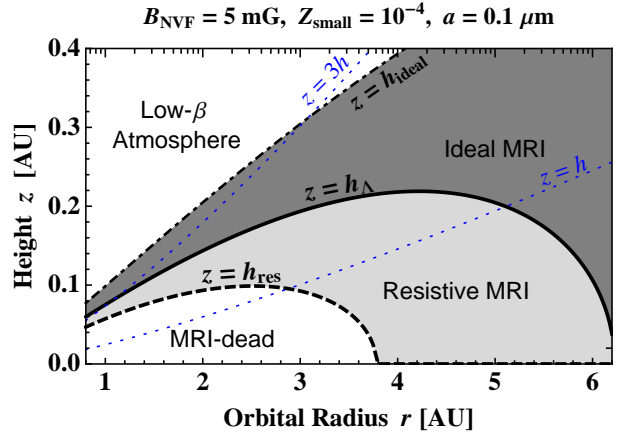}
\caption{Example of the dead zone size. 
The dot-dashed curve shows $z=h_\ideal$ for $B_\nvf=5~{\rm mG}$, 
while the solid and dashed curves show $z=h_\Lambda$ and $z=h_\res$, respectively,
for $B_\nvf=5~{\rm mG}$, $Z_{\rm small} = 10^{-4}$, and $a=0.1\micron$. 
For reference, $z=h$ and $3h$ are shown by the dotted curves. 
}
\label{fig:hr}
\end{figure}
Equations~\eqref{eq:dvmid} and \eqref{eq:pred_core} predict 
$\delta v_{g,\md}$ as a function of $h_\res$, $h_\Lambda$, and $B_\nvf$.
We calculate the critical heights by considering 
the ionization balance at each $z$ taking into account grain charging. 
We use the analytic solution of the ionization balance equations derived by \citet{O09},  
which gives the ionization degree for arbitrary dust size distribution.
The ionizing sources we include are
Galactic cosmic rays \citep{UN09}, stellar X-rays \citep{IG99,BG09}, 
stellar energetic protons \citep{TD09},
and radionuclides \citep{UN09}.
Inclusion of cosmic and stellar protons gives the minimum estimate of the dead zone size
since strong T-Tauri winds may shield these particles well above disk surfaces. 
From the ionization balance, we calculate 
the vertical profile of $\eta$ \citep{BB94}, and obtain
$h_\res$ and $h_\Lambda$ for given $B_\nvf$.
Figure~\ref{fig:hr} plots $z=h_\res$ and $z=h_\Lambda$ 
versus $r$ for $B_\nvf = 5~{\rm mG}$ 
assuming that $0.1~\micron$ sized grains are uniformly mixed in the gas 
with mass abundance $Z_{\rm small} \equiv \Sigma_{\rm small}/\Sigma_g = 10^{-4}$.
Note that the dead zone size depends on $B_\nvf$; 
the larger $B_\nvf$ is, the smaller $h_\res$ and $h_\Lambda$ are. 

\section{Icy Planetesimal Formation Across the Fragmentation Barrier}\label{sec:ice}
Planetesimal formation via direct coagulation is limited by the fact that 
marginally decoupled ($\Omega t_s \sim 1$, $a\sim1~{\rm m}$) dust aggregates 
experience high-speed collisions that can lead to catastrophic disruption (the so-called fragmentation barrier; \citealt{BDH08}).
If the disk is laminar, the maximum collision velocity 
is determined by the differential radial drift velocity 
$\sim |v_{dr,\max}| \sim 30~{\rm m~s^{-1}}$ \citep{BDH08}.
Recent numerical collision experiments \citep{W+09} show  
that aggregates made of submicron-sized icy grains 
stick at collision velocities up to $\approx 50~{\rm m~s^{-1}}$, 
suggesting that icy planetesimal formation via direct coagulation 
is possible in laminar disks.
However, if the disk has MRI-active layers, a random velocity of 
$\sim \delta v_{g,\md}$ is added to the collision velocity 
for $\Omega t_s \sim 1$ aggregates \citep{OC07}. 
The question is: can icy aggregates grow across the fragmentation barrier
even if the MRI-driven turbulence enhances the collision velocity?

\begin{figure}
\epsscale{1.}
\plotone{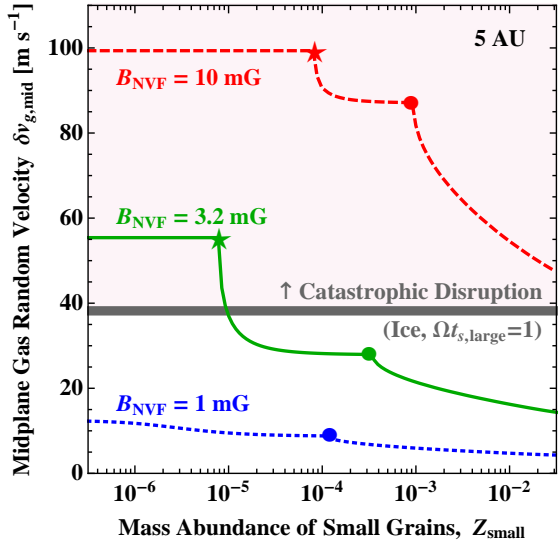}
\caption{Midplane gas random velocity $\delta v_{g,\md}$ at $5~\AU$
vs.~the mass abundance $Z_{\rm small}$ of small grains 
for different values of the NVF strength $B_\nvf$. 
The filled circles and stars mark $Z_{\rm small}$
below which $h_\res $ and $h_\Lambda$ vanish, respectively.
Above the thick gray line, the total collision velocity for
large ($\Omega t_{s,{\rm large}} \sim 1$) aggregates exceeds 
the catastrophic disruption threshold for ice \citep{W+09}.
 }
\label{fig:dvmid}
\end{figure}
To get a feeling of how the fragmentation barrier depends on dust size distribution 
and NVF, we begin with a simple two-population model in which large, marginally decoupled ($\Omega t_{s,{\rm large}} \sim 1$) aggregates coexist 
with $0.1~\micron$ sized small grains.
We calculate $\delta v_{g,\md}$ 
assuming that only the small grains contribute to the ionization balance. 
The mass abundance $Z_{\rm small}$ of the small grains is taken as a free parameter.
Figure~\ref{fig:dvmid} shows $\delta v_{g,\md}$ 
as a function of $Z_{\rm small}$ for different values of $B_\nvf$.
For fixed $B_\nvf$, $\delta v_{g,\md}$ increases with decreasing $Z_{\rm small}$
because the dead zone is smaller when small grains are less abundant.
The thick gray line shows $\delta v_{g,\md} = 38~{\rm m~s^{-1}}$; above this line, 
the total collision velocity for two $\Omega t_s \sim 1$ aggregates,
$\Delta v_{\rm large} \sim (\delta v_{g,\md}^2 + v_{dr,\max}^2)^{1/2}$, 
exceeds the disruption velocity $v_{\rm disr} = 50~{\rm m~s^{-1}}$ 
for icy aggregates \citep{W+09}.
If $B_\nvf = 1.1~{\rm mG}$, $\Delta v_{\rm large}$ falls below $v_{\rm disr}$
for all values of $Z_{\rm small}$.
The reason is two-fold. 
For such small $B_\nvf$, a substantially large ($h_\Lambda \sim h$) 
dead zone is present even in the absence of small grains 
(note that $h_\Lambda$ is larger when $B_\nvf$ is smaller).
In addition, the small $B_\nvf$ leads to a low saturation level in the upper active
layer, resulting in a low gas velocity dispersion 
(note that $\delta v_{g,\md} \propto B_\nvf$ when $h_\Lambda \gg h/4$).
However, if $B_\nvf$ goes up to $3.2~{\rm mG}$, 
$\Delta v_{\rm large}$ exceeds $v_{\rm disr}$
for $Z_{\rm small} \la 10^{-5}$, for which 
the active layer reaches the midplane ($h_\Lambda=0$).
For $B_\nvf \ga 10~{\rm mG}$, $\Delta v_{\rm large}$ exceeds $v_{\rm disr}$
even if $Z_{\rm small}$ is as large as the interstellar value $10^{-2}$. 

\begin{figure*}
\epsscale{1.}
\plotone{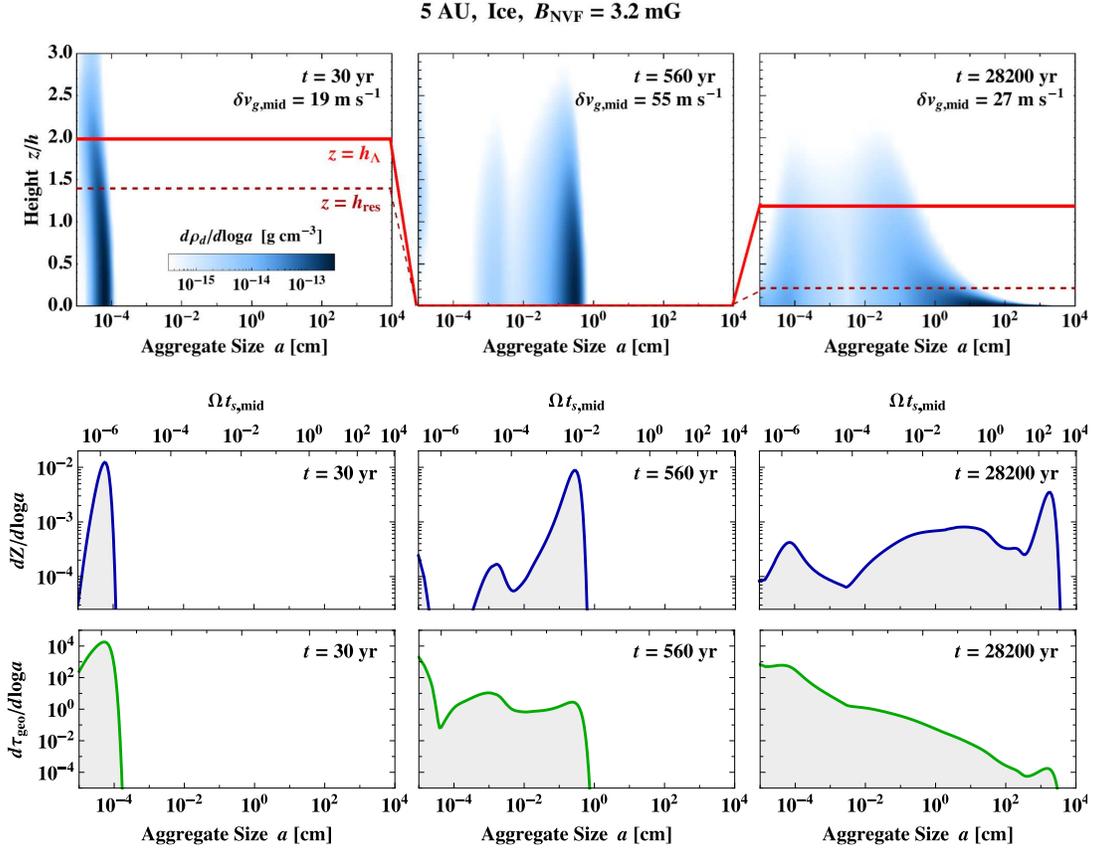}
\caption{Result of a coagulation simulation for icy dust at 5~AU for $B_\nvf = 3.2~{\rm mG}$.
The top density plots show the dust mass density $\rho_d$ 
per unit logarithmic particle radius $\log a$ at different heights $z$ and 
different times $t$. 
The solid and dashed lines in the top panels mark 
the dead zone critical heights $z = h_\Lambda$ and $z=h_\res$, respectively.
The middle and bottom panels show the mass abundance $Z$
and geometric optical depth $\tau_{\rm geo}$ of dust 
per unit $\log a$, respectively.  
Shown in the top of the middle panels is the dimensionless stopping time $\Omega t_{s,\md}$ of the dust particles at the midplane.
}
\label{fig:2D}
\end{figure*}
For $B_\nvf \sim 3~{\rm mG}$, large aggregates can grow across 
the fragmentation barrier only if small grains are sufficiently abundant.
To show that it is indeed possible, we simulate the evolution of full dust size distribution at 5 AU 
using a  coagulation--advection--diffusion equation \citep[e.g.,][]{C07}.
In this simulation, we begin with $0.1~\micron$ icy dust grains well mixed in the vertical direction, 
and follow the evolution of the vertical dust size distribution due to coagulation/fragmentation
and vertical settling/diffusion.  
The collision velocity $\Delta v$ of dust particles takes into account Brownian motion, 
systematic drift in the mean gas motion \citep{NSH86}, 
and turbulence-driven random motion \citep{OC07} that depends on $\delta v_g$. 
We take $\delta v_g$ to be time-dependent, namely, to 
be consistent with the dead zone size calculated 
from the full dust size distribution \citep{O09} at each time step.
The total dust mass abundance $Z = \Sigma_d/\Sigma_g$ is taken to be $10^{-2}$, 
and the loss of dust materials due to the radial drift is neglected.
Indeed, coagulation proceeds faster than radial migration 
if the dust grows into highly porous aggregates \citep{OTKW12}.
However, in order to make our model as simple as possible, 
we neglect both porosity evolution and radial migration.
The mass of an aggregate after collision is given by $m_t + sm_p$, 
where $m_t$ and $m_p(<m_t)$ are the masses of the target and projectile, 
and $s(\leq 1)$ is the dimensionless sticking efficiency that depends on $\Delta v$.
We assume $s=1$ for $\Delta v<10~{\rm m~s^{-1}}$ and 
$s=1-\ln(\Delta v/10~{\rm m~s^{-1}})/\ln 5$ for $\Delta v>10~{\rm m~s^{-1}}$
in accordance with the result of numerical collision experiments for icy aggregates \citep{W+09}.
The target loses its mass ($s<0$) when $\Delta v$ is larger than 
the disruption threshold $v_{\rm disr} = 50~{\rm m~s^{-1}}$.
The fragments (whose total mass is $(1-s)m_p$) 
are assumed to be in the form of $0.1~\micron$ sized constituent grains.
This assumption is quite simplistic, but still is reasonable 
as a first-order approximation 
since the total mass of fragments in aggregate--aggregate collision 
tends to be dominated by the smallest ones \citep{WPK05,W+09}.

Figure~\ref{fig:2D} shows the simulation result for $B_\nvf = 3.2~{\rm mG}$.
The top panels display the vertical dust mass density  
$\rho_d$ per unit logarithmic particle radius $\log a$
at different times $t$ after the initial state.
Initially, dust particles grow without catastrophic disruption because of low collision velocities.
As the particles grow, disruption becomes more and more  significant, 
because the collision velocity increases with increasing stopping time
and because the dead zone shrinks with the depletion of small grains.
The second effect becomes prominent when $t\approx560~{\rm yr}$, 
at which the dead zone disappears and consequently $\delta v_{g,\md}$ reaches $55~{\rm m~s^{-1}}$.
However, the large amount of tiny grains produced by the catastrophic disruption
are quickly diffused to high altitudes, reduce the ionization degree, 
and finally ``revive'' the dead zone. 
The revived dead zone suppresses $\delta v_{g,\md}$ down to $27~{\rm m~s^{-1}}$, 
which is low enough for larger aggregates to continue growing.
In this way, larger aggregates grow beyond the meter-size fragmentation barrier.

The middle panels of Figure~\ref{fig:2D} plot  
the dust mass abundance $Z$ per unit $\log a$, while 
the bottom panels show the vertical geometric optical depth $\tau_{\rm geo}$
of dust per unit $\log a$.
We see that  the dust mass is dominated by large aggregates 
while the optical depth is dominated by small fragments.
This indicates that small fragments carry a minor fraction of the total dust mass 
but still help the grow of larger aggregates by providing a large surface area 
needed to maintain a low enough ionization degree.

\section{Rocky Planetesimal Formation via Secular GI}\label{sec:rock}
Coagulation of rocky aggregates is severely restricted by the fragmentation barrier
since the disruption threshold is as low as $1$--$5~{\rm m~s^{-1}}$ \citep{W+09,G+10}.
Using that the radial drift speed 
$|v_{dr}|$ is approximated as $2|v_{dr,\max}|\Omega t_s$ 
for $\Omega t_s \ll 1$\citep{W77} and assuming that
$v_{\rm disr} \approx 5~{\rm m~s^{-1}}$ and 
$|v_{dr,\max}| \approx 30~{\rm m~s^{-1}}$,
 we find that  $|v_{dr}|$ 
reaches $v_{\rm disr}$ when $\Omega t_s \approx 0.08$, which corresponds to 
$a\sim 10~{\rm cm}$ at $\sim 1~\AU$.
Hence, a simple coagulation scenario does not 
account for rocky planetesimal formation even without turbulence.

One mechanism that can lead to rocky planetesimal formation is 
the so-called secular GI (\citealt*{Y11}). 
It is a gravitational collapse of dust materials driven by the combination of self-gravity and gas friction.
An important feature of the secular GI is that it works even if $\Omega t_s \ll 1$. 
Thus, the secular GI can allow gravitational collapse of dust particles 
whose growth is limited by the fragmentation barrier.
Instead, the secular GI requires sufficiently weak radial dust diffusion 
in order for the particles to collapse faster than they drift inward.
For $\Omega t_s \approx 0.1$, the radial diffusion coefficient $D_r$ must be lower than 
$10^{-5} c_s h$ at the midplane \citep{Y11,TI12}.

\begin{figure}[t]
\epsscale{1.}
\plotone{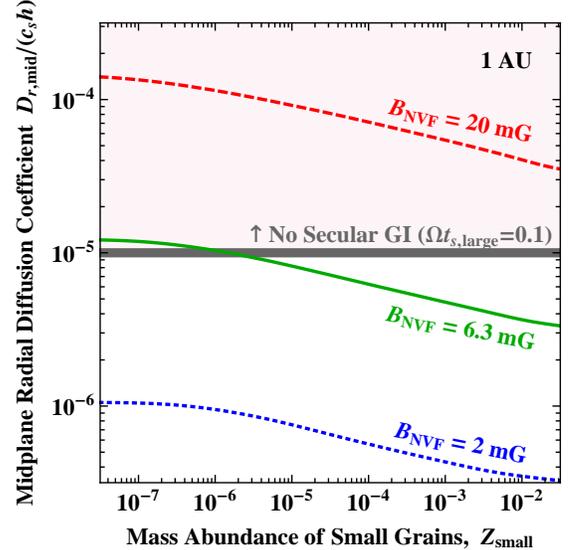}
\caption{Midplane radial diffusion coefficient $D_{r,\md}$ at 1~AU
vs.~the mass abundance $Z_{\rm small}$ of small grains 
for different values of the NVF strength $B_\nvf$. 
Below the thick gray line, the fragmentation-limited  rocky aggregates 
($\Omega t_{s,{\rm large}}\approx0.1$)
experience secular gravitational collapse faster than they drift  inward \citep{Y11,TI12}.
}
\label{fig:D}
\end{figure}

To assess whether the secular GI operates for the fragmentation-limited aggregates, 
we estimate the MRI-driven diffusion coefficient assuming that the aggregates
coexist with $0.1~\micron$ sized fragments of mass abundance $Z_{\rm small}$.
We also assume that $D_r \approx D_z$, which is $\approx D_{z0}$ 
for $\Omega t_s\approx0.1\ll1$ (see Equation~\eqref{eq:Dz0}).
Figure~\ref{fig:D} shows the midplane radial  diffusion coefficient $D_{r,\md}$ 
at 1~AU as a function of $Z_{\rm small}$ for different values of $B_\nvf$.
We find that $D_{r,\md}$ exceeds $10^{-5}c_sh$ for $B_\nvf\ga10~{\rm mG}$
even if $Z_{\rm small}$ is as large as the interstellar value $10^{-2}$.
Thus, the secular GI of the fragmentation-limited aggregates requires 
$B_\nvf\la10~{\rm mG}$.
This requirement is similar to to that for direct icy planetesimal formation shown 
in Section~\ref{sec:ice}.

\section{Conclusion and Discussion}
We have investigated how planetesimal formation depends 
on the amount of tiny grains and the strength of the NVF. 
For MMSN disks, we have shown that the existence of a large dead zone {\it and} 
an NVF weaker than 10~mG is preferable for planetesimal formation 
via both direct coagulation and secular GI.
The obtained criterion for the NVF depends on the disk surface density $\Sigma_g$, 
since $\delta v_{g,\md}\propto B_\nvf/\sqrt{\Sigma_g}$ in the presence of a large dead zone
(see Equations~\eqref{eq:dvmid} and \eqref{eq:pred_core}).
If $\Sigma_g$ is 10 times larger than 
the MMSN value, then the upper limit on $B_\nvf$ goes up to 30~mG.

We have neglected the effects of non-ohmic magnetic diffusivities. 
Ambipolar diffusion may stabilize MRI near the upper boundary of the active layer \citep{B11a}.
The effect of Hall diffusion is more uncertain; it can stabilize or destabilize MRI 
depending on the sign of NVF relative to the disk rotation axis \citep{WS12}.
Inclusion of these effects may change our results quantitatively, 
but the general trend that a weak NVF is preferable may be unchanged.

Weak turbulence is also beneficial to the growth of solid objects larger than planetesimals.
Density fluctuations in turbulence gravitationally interact with planetesimal-size objects
and thereby enhance their collision velocity.
If MRI is fully active, the resulting gravitational stirring 
likely causes catastrophic disruption of planetesimals \citep[e.g.,][]{IGM08,NG10}. 
However, \citet{GNT12} have recently shown that even weakly bound planetesimals 
are able to grow if a dead zone is present and 
if the NVF is weaker than $3~{\rm mG}$ (assuming the MMSN surface density).
Thus, a weak NVF and a wide dead zone are preferable for the growth of solid bodies 
up to protoplanets.

A realistic range of the NVF strength is poorly constrained by direct observations
since the differential rotation of the disk can produce toroidal magnetic fields 
as strong as  $0.1$--$1~{\rm G}$ even inside the dead zone \citep{TS08}.
An important property of NVF is that the total NVF is a conserved quantity of a magnetized disk.
Thus, the total NVF of a protoplanetary disk is directly determined by how the disk formed
from weakly magnetized ($\sim10~{\rm\mu G}$; \citealt*{HC05}) molecular clouds. 
Nevertheless, what fraction of the magnetic flux is brought to the planet-forming inner disk region
is not evident because non-ideal MHD processes also 
work during the disk formation \citep[e.g.,][]{MIM11}.
Furthermore, on longer timescales, the NVF may be radially transported due to 
inward mass accretion \citep{RL08} and/or outward macroscopic magnetic diffusion 
\citep{LPP94} in the active layers.
So far, the origin and global transport of NVF has not been paid attention in the context of 
planet(esimal) formation.
We hope that this Letter will encourage further investigation on this issue. 

\acknowledgments
We thank Shu-ichiro Inutsuka, Takeru Suzuki, Neal Turner, 
Masahiro Machida, Sanemichi Takahashi, Taku Takeuchi, Takayuki Muto, 
and Hidekazu Tanaka for useful discussion,
and the anonymous referee for prompt and useful comments. 
S.O. is supported by Grant-in-Aid for JSPS Fellows ($22\cdot 7006$) 
from MEXT of Japan.



\end{document}